\documentclass[preprint,prd,aps,showpacs,showkeys,nofootinbib]{revtex4}
\usepackage{graphicx}
\usepackage{dcolumn}
\usepackage{bm}
\begin{document}
\title{The study of lepton EDM in CP violating BLMSSM}
\author{Shu-Min Zhao$^1$\footnote{zhaosm@hbu.edu.cn}, Tai-Fu Feng$^{1}$\footnote{fengtf@hbu.edu.cn},
 Xi-Jie Zhan$^1$, Hai-Bin Zhang$^{1,2}$, Ben Yan$^{1}$}
\affiliation{$^1$ Department of Physics, Hebei University, Baoding 071002,
China\\
$^2$ Department of Physics, Dalian University of Technology, Dalian 116024, China}
\date{\today}
\begin{abstract}
In the supersymmetric model with local gauged baryon
and lepton numbers(BLMSSM), the CP violating effects are considered to study the lepton
electric dipole moment(EDM). The CP violating phases in BLMSSM are more than those in
the standard model(SM) and can give large contributions. The analysis of the EDMs for the leptons $e,\mu, \tau$ is shown
in this work. It is in favour of exploring the source of CP violation and probing the
physics beyond SM.

\end{abstract}


\maketitle

\section{Introduction}
\indent\indent
The theoretical predictions for EDMs of leptons and neutron are very
small in SM. The estimated SM value for the electron EDM is about $|d_{e}|\simeq10^{-38}e.cm$\cite{deSM}, which
is too small to be detected by the current experiments. The ACME Collaboration\cite{ACME} report the new result
of electron EDM $d_e=(-2.1\pm3.7_{stat}\pm2.5_{syst})\times10^{-29}e.cm$. The upper bound of electron EDM is
$|d_e|<8.7\times10^{-29}e.cm$ at the 90\% confidence level.
Therefore, if large EDM of electron is
probed, one can ensure it is the sinal of new physics beyond SM.
$|d_{\mu}|<1.9\times10^{-19}e.cm$ and $|d_\tau|<10^{-17}e.cm$ are the EDM upper bounds of $\mu$ and $\tau$
respectively\cite{EDMe1}. The minimal supersymmetric extension of SM
(MSSM) \cite{MSSM} is very attractive and
physicists have studied it for a long time. In MSSM, there are a lot of CP violating phases
and they can give large contributions to the EDMs of leptons and neutron.

In MSSM, when the CP violating phases are of normal size and the SUSY particles are at TeV scale,
big EDMs of elementary particles are obtained, and they can exceed the current experiment limits.
Three approaches are used to resolve
this problem. 1. make the CP violating phases small, i.e. $O(10^{-2})$. That is the so called tuning.
2. use mass suppression through making SUSY particles heavy(several TeV).
3. there is cancellation mechanism among the different components. For lepton EDM and neutron EDM,
the main parts of chargino and the neutralino contributions are cancelled\cite{CPxiangxiao}.

BLMSSM is the minimal supersymmetric extension of
the SM with local gauged B and L\cite{BLMSSM}. Therefore, it can explain both the asymmetry of matter-antimatter in the universe
and the data from neutrino oscillation experiment. We consider that BLMSSM is a favorite model beyond MSSM.
Extending SM, the authors study the model with B and L as
spontaneously broken gauge symmetries around ${\rm TeV}$ scale\cite{BLMSSM1}. The lightest CP-even Higgs mass and  the
 decays  $h^0\rightarrow\gamma\gamma$, $h^0\rightarrow ZZ (WW)$ are also studied in this model\cite{weBLMSSM}. In our
 previous works\cite{smneutron,sunfei}, we study the neutron EDM and $B^0-\bar{B}^0$ mixing in CP violating BLMSSM.

Research the MDMs \cite{g2} and EDMs\cite{EDMe,EDM} of leptons are the effective ways to probe new physics beyond the SM.
In MSSM, the one-loop contributions to lepton MDM and EDM are well studied, and
some two loop corrections are also investigated.
In the two Higgs doublet models with CP violation, the authors obtain the one-loop and Barr-Zee type two-loop contributions to fermionic
EDMs. A model-independent
study of $d_e$ in the SM is carried out\cite{xghe}. They take into account the right-handed neutrinos,
the neutrino see-saw mechanism and the framework of minimal flavor violation. Their results show that when neutrinos are
Majorana particles, $d_e$ can reach its experiment upper bound.

After this introduction, in section 2 we briefly introduce the main ingredients
of the BLMSSM.  The one-loop corrections to the lepton EDM are collected in
section 3. Section 4 is devoted to the numerical analysis for the dependence of lepton EDM on
the BLMSSM parameters.  We show our discussion and conclusion in section 5.

\section{The BLMSSM}
The local gauge group of BLMSSM\cite{BLMSSM} is $SU(3)_{C}\otimes SU(2)_{L}\otimes U(1)_{Y}\otimes U(1)_{B}\otimes U(1)_{L}$,
 where the exotic leptons are introduced to cancel $L$ anomaly. Similarly, they
introduce the exotic quarks to cancel the $B$ anomaly. In this work, the quarks, exotic quarks and
exotic leptons have none one-loop contribution to lepton EDM, so we do not introduce them in detail.
The Higgs mechanism is of solid foundation, because of the detection of the lightest CP even Higgs $h^0$ at LHC\cite{Higgs}.
The Higgs superfields are used to break lepton number spontaneously, and
 they need nonzero vacuum expectation values (VEVs).

 The superpotential of BLMSSM is shown as
\begin{eqnarray}
&&{\cal W}_{{BLMSSM}}={\cal W}_{{MSSM}}+{\cal W}_{B}+{\cal W}_{L}+{\cal W}_{X}\;.
\label{superpotential1}
\end{eqnarray}
Here, ${\cal W}_{{MSSM}}$ represents the superpotential of the MSSM. The concrete forms of
 ${\cal W}_{B},~{\cal W}_{L}$ and ${\cal W}_{X}$ are shown in the work\cite{weBLMSSM}.
 ${\cal W}_{L}$ includes the needed new term ${\cal W}_{L}(n)$, which is collected here
\begin{eqnarray}
&&{\cal W}_{L}(n)=Y_{\nu}\hat{L}\hat{H}_{u}\hat{N}^c+\lambda_{{N^c}}\hat{N}^c\hat{N}^c\hat{\varphi}_{L}
+\mu_{L}\hat{\Phi}_{L}\hat{\varphi}_{L}\;.
\label{superpotential-BL}
\end{eqnarray}

In BLMSSM, the complete soft breaking terms are very complex\cite{weBLMSSM,smneutron}, and only the terms
 relating with lepton are necessary for our study
\begin{eqnarray}
&&{\cal L}_{{soft}}(L)=-m_{{\tilde{N}^c}}^2\tilde{N}^{c*}\tilde{N}^c
-m_{{\Phi_{L}}}^2\Phi_{L}^*\Phi_{L}
-m_{{\varphi_{L}}}^2\varphi_{L}^*\varphi_{L}-\Big(M_{L}\lambda_{L}\lambda_{L}+h.c.\Big)
\nonumber\\
&&\hspace{2.0cm}
+\Big(A_{N}Y_{\nu}\tilde{L}H_{u}\tilde{N}^c+A_{{N^c}}\lambda_{{N^c}}\tilde{N}^c\tilde{N}^c\varphi_{L}
+B_{L}\mu_{L}\Phi_{L}\varphi_{L}+h.c.\Big).
\label{soft-breaking}
\end{eqnarray}

In order to break the local gauge symmetry $SU(2)_{L}\otimes U(1)_{Y}\otimes U(1)_{B}\otimes U(1)_{L}$
  down to the electromagnetic symmetry $U(1)_{e}$, the $SU(2)_L$ doublets $H_{u}$ and $H_{d}$ should obtain nonzero VEVs
$\upsilon_{u}$ and $\upsilon_{d}$. While the $SU(2)_L$ singlets $\Phi_{L}$ and $\varphi_{L}$ should obtain nonzero VEVs $\upsilon_{L}$ and $\overline{\upsilon}_{L}$ respectively. The needed Higgs fields and Higgs superfields are defined as
\begin{eqnarray}
&&H_{u}=\left(\begin{array}{c}H_{u}^+\\{1\over\sqrt{2}}\Big(\upsilon_{u}+H_{u}^0+iP_{u}^0\Big)\end{array}\right)\;,~~~~
H_{d}=\left(\begin{array}{c}{1\over\sqrt{2}}\Big(\upsilon_{d}+H_{d}^0+iP_{d}^0\Big)\\H_{d}^-\end{array}\right)\;,
\nonumber\\
&&\Phi_{L}={1\over\sqrt{2}}\Big(\upsilon_{L}+\Phi_{L}^0+iP_{L}^0\Big)\;,~~~~~~~~~~
\varphi_{L}={1\over\sqrt{2}}\Big(\overline{\upsilon}_{L}+\varphi_{L}^0+i\overline{P}_{L}^0\Big)\;.
\label{VEVs}
\end{eqnarray}

The detailed discussion of Higgs mass matrices can be found in
Ref.\cite{weBLMSSM}. The super field $\hat{N}^c$ in BLMSSM leads to that the neutrinos and sneutrinos are doubled as those in MSSM.
Through the see-saw mechanism, light neutrinos obtain tiny masses.

In BLMSSM, there are 10 neutralinos: 4 MSSM neutralinos, 3 baryon neutralinos and 3
lepton neutralinos. The MSSM neutralinos, baryon neutralinos and lepton neutralinos do not mix with each other.
Baryon neutralinos  have zero
 contribution to the lepton EDM at one-loop level. While,
 lepton neutralinos can give contributions to lepton EDM through
  lepton-slepton-lepton neutralino coupling.
 The three lepton neutralinos are made up of $\lambda_L$
(the superpartners of the new lepton boson) and $\psi_{\Phi_L},\psi_{\varphi_L}$
 (the superpartners of the $SU(2)_L$ singlets $\Phi_L,\varphi_L$). Here, we show the mass term of lepton neutralinos.

\begin{equation}
\mathcal{L}_{\chi_L^0}=\frac{1}{2}(i\lambda_L,\psi_{\Phi_L},\psi_{\varphi_L})\left(     \begin{array}{ccc}
  2M_L &2v_Lg_L &-2\bar{v}_Lg_L\\
   2v_Lg_L & 0 &-\mu_L\\-2\bar{v}_Lg_L&-\mu_L &0
    \end{array}\right)  \left( \begin{array}{c}
 i\lambda_L \\ \psi_{\Phi_L}\\\psi_{\varphi_L}
    \end{array}\right)+h.c.\label{LN}
   \end{equation}
Three lepton neutralino masses are obtained from diagonalizing the mass mixing matrix in Eq.(\ref{LN}) by $Z_{N_L}$.

Though in BLMSSM there are six sleptons, their mass squared matrix is
different from that in MSSM, because of the contributions from Eqs.(\ref{superpotential-BL},\ref{soft-breaking}).
We deduce the corrected  mass squared matrix of slepton, and the matrix $Z_{\tilde{L}}$ is used to
diagonalize it
\begin{eqnarray}
&&\left(\begin{array}{cc}
  (\mathcal{M}^2_L)_{LL}&(\mathcal{M}^2_L)_{LR} \\
   (\mathcal{M}^2_L)_{LR}^{\dag} & (\mathcal{M}^2_L)_{RR}
    \end{array}\right).
\end{eqnarray}
$(\mathcal{M}^2_L)_{LL},~(\mathcal{M}^2_L)_{LR}$ and $(\mathcal{M}^2_L)_{RR}$ are shown as
\begin{eqnarray}
 &&(\mathcal{M}^2_L)_{LL}=\frac{(g_1^2-g_2^2)(v_d^2-v_u^2)}{8}\delta_{IJ} +g_L^2(\bar{v}_L^2-v_L^2)\delta_{IJ}
 +m_{l^I}^2\delta_{IJ}+(m^2_{\tilde{L}})_{IJ},\nonumber\\&&
 (\mathcal{M}^2_L)_{LR}=\frac{\mu^*v_u}{\sqrt{2}}(Y_l)_{IJ}-\frac{v_u}{\sqrt{2}}(A'_l)_{IJ}+\frac{v_d}{\sqrt{2}}(A_l)_{IJ},
 \nonumber\\&& (\mathcal{M}^2_L)_{RR}=\frac{g_1^2(v_u^2-v_d^2)}{4}\delta_{IJ}-g_L^2(\bar{v}_L^2-v_L^2)\delta_{IJ}
 +m_{l^I}^2\delta_{IJ}+(m^2_{\tilde{R}})_{IJ}.
\end{eqnarray}

The super field $\hat{N}^c$ is introduced in BLMSSM, so the neutrino mass
matrix and the sneutrino mass squared matrix are both $6\times 6$ and more complicated than those in MSSM.
 In the left-handed basis $(\nu,N^c)$, we deduce the mass
matrix of neutrino after symmetry breaking
    \begin{eqnarray}
-\mathcal{L}_{mass}^{\nu}=(\bar{\nu}^I_R,\bar{N}^{cI}_R )\left(\begin{array}{cc}
  0&\frac{v_u}{\sqrt{2}}(Y_{\nu})_{IJ} \\
   \frac{v_u}{\sqrt{2}}(Y^{T}_{\nu})_{IJ}  & \frac{\bar{v}_L}{\sqrt{2}}(\lambda_{N^c})_{IJ}
    \end{array}\right) \left(\begin{array}{c}
   \nu^J_L\\   N^{cJ}_L
    \end{array}\right)+h.c.
      \end{eqnarray}
      With the unitary transformations
      \begin{eqnarray}
&&\left(\begin{array}{l}\nu_{1L}^I\\\nu_{2L}^I\end{array}\right)
=U_{\nu^{IJ}}^\dagger\left(\begin{array}{c}
   \nu^J_L\\   N^{cJ}_L
    \end{array}\right)\;,\;\;
\left(\begin{array}{l}\nu_{1R}^I\\\nu_{2R}^I\end{array}\right)
=W_{\nu^{IJ}}^\dagger\left(\begin{array}{c}
   \nu^J_R\\ N^{cJ}_R
    \end{array}\right),
\end{eqnarray}
the mass matrix of neutrino is diagonalized as
      \begin{eqnarray}
W_{\nu^{IJ}}^{\dag}\left(\begin{array}{cc}
  0&\frac{v_u}{\sqrt{2}}(Y_{\nu})_{IJ} \\
   \frac{v_u}{\sqrt{2}}(Y^{T}_{\nu})_{IJ}  & \frac{\bar{v}_L}{\sqrt{2}}(\lambda_{N^c})_{IJ}
    \end{array}\right) U_{\nu^{IJ}}=\texttt{diag}(m_{\nu_1^I},m_{\nu_2^I}).
      \end{eqnarray}

The trilinear sneutrino-Higgs-sneutrino coupling $A_{{N^c}}\lambda_{{N^c}}\tilde{N}^c\tilde{N}^c\varphi_{L}$ in the
soft breaking terms $\mathcal{L}_{soft}(L)$ leads to large sneutrino masses.
The VEV of $\varphi_{L}$ is $\frac{1}{\sqrt{2}}\bar{v}_L$,
and the contribution from this term to sneutrino masses is at the order of $A_{{N^c}}\lambda_{{N^c}}\bar{v}_L$.
The super potential of BLMSSM includes the new term ${\cal W}_{L}(n)$.
Then two functions and the scalar supersymmetric potential are shown here
\begin{eqnarray}
&&F_i=\frac{\partial W}{\partial A_i}, ~~~~~~~~D^a=gA^*_iT^a_{ij}A_j,
\nonumber\\&&
V=\frac{1}{2}D^aD^a+F_i^*F_i,
\end{eqnarray}
where $A_i$ represent the scalar fields.
The first term $Y_{\nu}\hat{L}\hat{H}_{u}\hat{N}^c$ in ${\cal W}_{L}(n)$ is suppressed by $Y_\nu$.
Using the formula Eq.(2), the second term $\lambda_{{N^c}}\hat{N}^c\hat{N}^c\hat{\varphi}_{L}$ in ${\cal W}_{L}(n)$ can give
important contributions to large sneutrino masses through nonzero VEV of Higgs superfield $\varphi_{L}$. This type correction is at the order of $\lambda^*_{{N^c}}\lambda_{{N^c}}\bar{v}_L^2$.
The orders of both dominant contributions respectively are $A_{{N^c}}\lambda_{{N^c}}\bar{v}_L$ and $\lambda^*_{{N^c}}\lambda_{{N^c}}\bar{v}_L^2$,
that are much larger than the product of neutrino Yukawa and SUSY
scale.
The mass squared  matrix of the sneutrino is obtained
 from the superpotential and the soft breaking terms from Eqs.(\ref{superpotential-BL})(\ref{soft-breaking}),
 \begin{eqnarray}
&&-\mathcal{L} _{\tilde{n}}^{mass}=\tilde{n}^{\dagger}\cdot
{\cal M}_{\tilde{n}}^2\cdot\tilde{n}
\label{SQmass-2/3},
\end{eqnarray}
 with $\tilde{n}^{T}=(\tilde{\nu},\tilde{N}^{c*})$.
 The sneutrinos are enlarged by the superfield $\hat{N}^c$ and the mass squared matrix of sneutrino reads as
    \begin{eqnarray}
  && {\cal M}^2_{\tilde{n}}(\tilde{\nu}_{I}^*\tilde{\nu}_{J})=\frac{g_1^2+g_2^2}{8}(v_d^2-v_u^2)\delta_{IJ}+g_L^2(\overline{v}^2_L-v^2_L)\delta_{IJ}
   +\frac{v_u^2}{2}(Y^\dag_{\nu}Y_\nu)_{IJ}+(m^2_{\tilde{L}})_{IJ},\nonumber\\&&
   {\cal M}^2_{\tilde{n}}(\tilde{N}_I^{c*}\tilde{N}_J^c)=-g_L^2(\overline{v}^2_L-v^2_L)\delta_{IJ}
   +\frac{v_u^2}{2}(Y^\dag_{\nu}Y_\nu)_{IJ}+2\overline{v}^2_L(\lambda_{N^c}^\dag\lambda_{N^c})_{IJ}\nonumber\\&&
   \hspace{1.8cm}+(m^2_{\tilde{N}^c})_{IJ}+\mu_L\frac{v_L}{\sqrt{2}}(\lambda_{N^c})_{IJ}
   -\frac{\overline{v}_L}{\sqrt{2}}(A_{N^c})_{IJ}(\lambda_{N^c})_{IJ},\nonumber\\&&
   {\cal M}^2_{\tilde{n}}(\tilde{\nu}_I\tilde{N}_J^c)=\mu^*\frac{v_d}{\sqrt{2}}(Y_{\nu})_{IJ}-v_u\overline{v}_L(Y_{\nu}^\dag\lambda_{N^c})_{IJ}
   +\frac{v_u}{\sqrt{2}}(A_{N})_{IJ}(Y_\nu)_{IJ}.
   \end{eqnarray}
Using the formula
$Z_{\nu^{IJ}}^\dag{\cal M}_{\tilde{n}}^2Z_{\nu^{IJ}}=\texttt{diag}(m_{\tilde{\nu}^{1}_1}^2,m_{\tilde{\nu}^{2}_1}^2,m_{\tilde{\nu}^{3}_1}^2
,m_{\tilde{\nu}^{1}_2}^2,m_{\tilde{\nu}^{2}_2}^2,m_{\tilde{\nu}^{3}_2}^2)$, the masses of the sneutrinos are obtained.

Because of the introduction of the  superfield $\hat{N}^c$ in BLMSSM, 
the corrected chargino-lepton-sneutrino coupling is adapted as
\begin{eqnarray}
&&\mathcal{L}_{\chi^{\pm}L\tilde{\nu}}=-\sum_{I,J=1}^3\sum_{i,j=1}^2\bar{\chi}^-_j
\Big((Y_l)_{IJ} Z_-^{2j*}(Z_{\nu^{IJ}}^\dag)^{i1}\omega_+\nonumber\\&&+
[\frac{e}{s_W}Z_+^{1j}(Z_{\nu^{IJ}}^\dag)^{i1}+(Y_\nu)_{IJ}Z_+^{2j}(Z_{\nu^{IJ}}^\dag)^{i2}]\omega_-
\Big)e^J\tilde{\nu}^{I*}_i+h.c.,
\end{eqnarray}
with $\omega_{\mp}=\frac{1\mp\gamma_5}{2}$.
Here we use the abbreviated form, $s_W=\sin\theta_W,~c_W=\cos\theta_W$, and $\theta_W$ is the Weinberg angle.

From the interactions of gauge and matter multiplets
$ig\sqrt{2}T^a_{ij}(\lambda^a\psi_jA_i^*-\bar{\lambda}^a\bar{\psi}_iA_j)$,
the lepton-slepton-lepton neutralino couplings are deduced here
\begin{eqnarray}
&&\mathcal{L}_{l\chi_L^0\tilde{L}}=\sqrt{2}g_L\bar{\chi}_{L_j}^0\Big(Z_{N_L}^{1j}Z_{L}^{Ii}\omega_-
+Z_{N_L}^{1j*}Z_{L}^{(I+3)i}\omega_+\Big)l^I\tilde{L}_i^++h.c.
\end{eqnarray}
This type couplings give new contributions beyond MSSM to lepton EDM.
Compared with MSSM, there are four new CP violating sources:
1. the mass $M_L$ of gaugino $\lambda_L$;
2. the superfield Higgsino's mass $\mu_L$, which is included in the mass matrices of both sneutrino and lepton neutralino.
3. $v_L$ in $\Phi_L$; 4.$\bar{v}_L$ in $\varphi_L$. In general, we take $v_L$ and $\bar{v}_L$ as real parameters to
simplify the numerical discussion.

\section{formulation}
To obtain the lepton EDM,
we use the effective Lagrangian method, and the Feynman amplitudes can be expressed by these dimension-6 operators.
\begin{eqnarray}
&&\mathcal{O}_1^{\mp}=\frac{1}{(4\pi)^2}\bar{l}(i\mathcal{D}\!\!\!\slash)^3\omega_{\mp}l,
\nonumber\\
&&\mathcal{O}_2^{\mp}=\frac{eQ_f}{(4\pi)^2}\overline{(i\mathcal{D}_{\mu}l)}\gamma^{\mu}
F\cdot\sigma\omega_{\mp}l,
\nonumber\\
&&\mathcal{O}_3^{\mp}=\frac{eQ_f}{(4\pi)^2}\bar{l}F\cdot\sigma\gamma^{\mu}
\omega_{\mp}(i\mathcal{D}_{\mu}l),\nonumber\\
&&\mathcal{O}_4^{\mp}=\frac{eQ_f}{(4\pi)^2}\bar{l}(\partial^{\mu}F_{\mu\nu})\gamma^{\nu}
\omega_{\mp}l,\nonumber\\&&
\mathcal{O}_5^{\mp}=\frac{m_l}{(4\pi)^2}\bar{l}(i\mathcal{D}\!\!\!\slash)^2\omega_{\mp}l,
\nonumber\\&&\mathcal{O}_6^{\mp}=\frac{eQ_fm_l}{(4\pi)^2}\bar{l}F\cdot\sigma
\omega_{\mp}l.
\end{eqnarray}

with $\mathcal{D}_{\mu}=\partial_{\mu}+ieA_{\mu}$,  $l$ denoting the lepton
fermion, $m_{l}$ being the lepton mass, $F_{\mu\nu}$ being the electromagnetic field strength.
Adopting on-shell condition for external leptons, only
$\mathcal{O}_{2,3,6}^{\mp}$ contribute to lepton EDM. Therefore, the Wilson coefficients of the operators $\mathcal{O}_{2,3,6}^{\mp}$ in the effective Lagrangian are of interest.

 The lepton EDM is expressed as \begin{eqnarray}
&&{\cal L}_{{EDM}}=-{i\over2}d_{l}\overline{l}\sigma^{\mu\nu}\gamma_5
lF_{{\mu\nu}}
\label{eq1}.
\end{eqnarray}
The fermion EDM
is a CP violating amplitude which can not be obtained at tree level in the fundamental interactions.
However, in the CP violating electroweak theory, one loop diagrams should contribute nonzero value to fermion EDM. Considering
the relations between the Wilson coefficients $C_{2,3,6}^{\mp}$ of the operators $\mathcal{O}_{2,3,6}^{\mp}$\cite{EDM}, the lepton EDM
is obtained
\begin{eqnarray}
d_l=-\frac{2eQ_fm_l}{(4\pi)^2}\texttt{Im}(C_2^++C_2^{-*}+C_6^+).
\end{eqnarray}
The one-loop triangle diagrams in BLMSSM are divided into three types according to the virtual particles:
1 the neutralino-slepton diagram; 2 the
chargino-sneutrino diagram; 3 the lepton neutralino-slepton diagram.
 After the calculation, using the on-shell condition for
 the external leptons, we obtain the one-loop
contributions to lepton EDM.
\begin{eqnarray}
&&d_{l^I}=
\frac{e}{32\pi^2\Lambda_{NP}}\texttt{Im}\Big\{\sum_{i=1}^6\sum_{j=1}^4(\mathcal{A}_1)^I_{ij}
(\mathcal{A}_2)^I_{ij}\sqrt{x_{\chi^0_j}}\Big[\frac{\partial^2}{\partial x^2_{\tilde{L}_i}}\varrho_{2,1}(x_{\chi^0_j},x_{\tilde{L}_i})-
2\frac{\partial}{\partial x_{\tilde{L}_i}}\varrho_{1,1}(x_{\chi^0_j},x_{\tilde{L}_i})
\Big]\nonumber\\&&
+\sum_{i=1}^6\sum_{j=1}^32g_L^2(Z_{N_L}^{1j*})^2Z_{L}^{Ii*}Z_{L}^{(I+3)i}\sqrt{x_{\chi^0_{L_j}}}\Big[\frac{\partial^2}{\partial x^2_{\tilde{L}_i}}\varrho_{2,1}(x_{\chi^0_{L_j}},x_{\tilde{L}_i})-
2\frac{\partial}{\partial x_{\tilde{L}_i}}\varrho_{1,1}(x_{\chi^0_{L_j}},x_{\tilde{L}_i})
\Big]
\nonumber\\&&+\sum_{J=1}^3\sum_{i,j=1}^2(\mathcal{B}_1)^{IJ}_{ij}(\mathcal{B}_2)^{IJ}_{ij}\sqrt{x_{\chi^{\mp}_j}}\Big[
\frac{\partial^2}{\partial x^2_{\tilde{\nu}^J_i}}\varrho_{2,1}(x_{\chi^{\mp}_j},x_{\tilde{\nu}^J_i})-
2\frac{\partial}{\partial x_{\tilde{\nu}^J_i}}\varrho_{1,1}(x_{\chi^{\mp}_j},x_{\tilde{\nu}^J_i})\nonumber\\&&+2\frac{\partial}{\partial x_{\chi^{\mp}_j}}\varrho_{1,1}(x_{\chi^{\mp}_j},x_{\tilde{\nu}^J_i})
\Big]\Big\},
\end{eqnarray}
with $x_i$ denoting $\frac{m_i^2}{\Lambda^2_{NP}}$, $\Lambda_{NP}$ representing energy scale of the new physics.
The concrete form of the function $\varrho_{i,j}(x,y)$ is shown here
\begin{eqnarray}
\varrho_{i,j}(x,y)=\frac{x^i\ln^jx-y^i\ln^jy}{x-y}.
\end{eqnarray}
The couplings $(\mathcal{A}_1)^I_{ij},(\mathcal{A}_2)^I_{ij},(\mathcal{B}_1)^{IJ}_{ij}$ and $(\mathcal{B}_2)^{IJ}_{ij}$ read as
\begin{eqnarray}
&&(\mathcal{A}_1)^I_{ij}=\frac{e}{\sqrt{2}s_Wc_W}Z_{\tilde{L}}^{Ii*}(Z_N^{1j*}s_W+Z_N^{2j*}c_W)+
(Y_{l})^*_{II}Z_{\tilde{L}}^{(I+3)i*}Z_N^{3j*},\nonumber\\&&
(\mathcal{A}_2)^I_{ij}=-\frac{e\sqrt{2}}{c_W}Z_{\tilde{L}}^{(I+3)i}Z_N^{1j*}+
(Y_{l})_{II}Z_{\tilde{L}}^{Ii}Z_N^{3j*},\nonumber\\
&&(\mathcal{B}_1)^{IJ}_{ij}=
\frac{e}{s_W}Z_+^{1j*}(Z_{\nu^{IJ}})^{1i}+(Y_\nu)_{IJ}^*Z_+^{2j*}(Z_{\nu^{IJ}})^{2i},
\nonumber\\&&(\mathcal{B}_2)^{IJ}_{ij}=(Y_l)_{IJ} Z_-^{2j*}(Z_{\nu^{IJ}})^{1i*}.
\end{eqnarray}
The matrices $Z_{\tilde{L}},Z_N$ respectively diagonalize the mass matrices of slepton and neutralino.

To explicit the phase of $\lambda_L$ obviously in the one loop contributions, we suppose $M_L\gg\mu_L,g_Lv_L,g_L\bar{v}_L$.
Then the lepton netralino-slepton contributions are simplified as
\begin{eqnarray}
&&d_{l^I}^{\lambda_L\tilde{L}}=\frac{e}{8\pi^2\Lambda_{NP}^2}\texttt{Im}\Big\{\sum_{i=1}^6g_L^2Z_{L}^{Ii*}Z_{L}^{(I+3)i}
|M_L|e^{i\theta_L}\nonumber\\&&\times\Big[\frac{\partial^2}{\partial x^2_{\tilde{L}_i}}\varrho_{2,1}(\frac{4|M_L|^2}{\Lambda_{NP}^2},x_{\tilde{L}_i})-
2\frac{\partial}{\partial x_{\tilde{L}_i}}\varrho_{1,1}(\frac{4|M_L|^2}{\Lambda_{NP}^2},x_{\tilde{L}_i})
\Big]\Big\}.
\end{eqnarray}
In this formula, the CP violating phase $\theta_L$ is conspicuous.
\section{the numerical results}
For the numerical discussion, we take into account of the lightest neutral CP-even Higgs mass
$m_{_{h^0}}\simeq125.7\;{\rm GeV}$ \cite{Higgs} and the neutrino experiment data\cite{neutrino,neutrino1}
\begin{eqnarray}
&&\sin^22\theta_{13}=0.090\pm 0.009,~~\sin^2\theta_{12} =0.306_{-0.015}^{+0.018},
~~\sin^2\theta_{23}=0.42_{-0.03}^{+0.08},\nonumber\\
&&\Delta m_{\odot}^2 =7.58_{-0.26}^{+0.22}\times 10^{-5} {\rm eV}^2,
~~|\Delta m_{A}^2| =2.35_{-0.09}^{+0.12}\times 10^{-3} {\rm eV}^2.
\label{neu-oscillations2}
\end{eqnarray}
In our previous works, the neutron EDM and muon MDM are studied\cite{smneutron,g2BL}, so the constraints from them are also considered here.

 We give out the used parameters\cite{14pdg,g2BL,BiaoChen}
 \begin{eqnarray}
 &&m_e=0.51\times10^{-3}{\rm GeV},~m_{\mu}=0.105{\rm GeV},~m_{\tau}=1.777{\rm GeV},
\nonumber\\&&m_W=80.385{\rm GeV},~\alpha(m_Z)=1/128,~s_W^2(m_Z)=0.23,
\nonumber\\
&&\tan\beta_{L}=2,~L_4={3\over2},~m_{Z_L}=1{\rm TeV},~ \Lambda_{NP}=1000{\rm GeV},
\nonumber\\&&
(Y_{\nu})_{11}=1.3031*10^{-6},~(Y_{\nu})_{12}=9.0884*10^{-8},~(Y_{\nu})_{13}=6.9408*10^{-8},\nonumber\\&&
(Y_{\nu})_{22}=1.6002*10^{-6},~
(Y_{\nu})_{23}=3.4872*10^{-7},~(Y_{\nu})_{33}=1.7208*10^{-6},\nonumber\\&&\lambda_{N^c}=1,~g_L=1/6,
~~(A_{N^c})_{ii}=(A_{N})_{ii}=3000{\rm GeV},~~\texttt{for}~ i=1,2,3,\nonumber\\
&&m_{1}=M1*e^{i*\theta_1},~m_2=M2*e^{i*\theta_2},~\mu_H=MU*e^{i*\theta_{\mu}},~
\sqrt{\upsilon_{L}^2+\overline{\upsilon}_{L}^2}=\upsilon_{L_t}.
\end{eqnarray}

Here, $\theta_1,\theta_2$ and $\theta_{\mu}$ are the CP violating phases of the parameters $m_1,m_2$, and $\mu_H$.
We consider two new CP violating parameters with the phases $\theta_{\mu_L}$ and $\theta_{L}$
\begin{eqnarray}
\mu_L=muL*e^{i*\theta_{\mu_L}},~~~~ M_L=ML*e^{i*\theta_{L}}.
\end{eqnarray}
To simplify the numerical discussion, we use the following relations.
 \begin{eqnarray}
&&(m_{\tilde{L}}^2)_{11}=(m_{\tilde{L}}^2)_{22}=(m_{\tilde{L}}^2)_{33}=S_L,~~
~(m_{\tilde{R}}^2)_{11}=(m_{\tilde{R}}^2)_{22}=(m_{\tilde{R}}^2)_{33}=S_R,\nonumber\\&&
(m^2_{\tilde{N}^c})_{11}=(m^2_{\tilde{N}^c})_{22}=(m^2_{\tilde{N}^c})_{33}=S_\nu, ~~~~~
(A_l)_{11}=(A_l)_{22}=(A_l)_{33}=AL.
\end{eqnarray}
If we do not specially declare, the non-diagonal elements of the used parameters should be zero.

To study the effects to lepton EDM from the non-diagonal elements of the used parameters, we consider
the constraints from the lepton flavor violating processes $l_j\to l_i+\gamma$ and
$l_j\to 3l_i$. The experiment upper bounds of $Br(\mu \to e+\gamma)$ and
$Br(\mu \to 3e)$ are respectively $5.7\times10^{-13}$ and $1.0\times10^{-12}$.
They are both strict and set severe limits on the parameter space, especially for the sensitive
parameters including non-diagonal elements for the lepton flavor violation.
In our prepared work\cite{BLLFV}, we study $Br(\mu \to e+\gamma)$ and $Br(\mu \to 3e)$, and find that the virtual particle masses,
$\tan\beta$ and the ratios of non-diagonal elements to diagonal elements for the slepton(sneurino)
 mass squared matrices are important parameters. When the slepton and sneutrino are at TeV order, $\tan\beta$ should
be in the region $10\sim 20$. The effects to $\mu \to e+\gamma$ and $\mu \to 3e$ from
the non-diagonal elements of $m^2_{\tilde{N^c}},A_{N^c}$ and $A_N$ are not large. On the other hand,
the non-diagonal elements of $m^2_{\tilde{L}},m^2_{\tilde{R}}$ influence the both LFV processes strong.
With the supposition $(m_{\tilde{L}}^2)_{ij}=(m_{\tilde{R}}^2)_{ij}=FL^2$, $FL$ should be in the range $(0\sim 500){\rm GeV}$ except
extreme parameter space.

\subsection{the electron EDM}
At first, we study electron EDM, because its upper bound is the most strict one.
The CP violating phases $\theta_1,\theta_2,\theta_{\mu},\theta_{\mu_L},\theta_L$, and
other parameters have close relationships with electron EDM. In this subsection, we suppose
$S_\nu=1.0\times10^6{\rm GeV}^2,m_L=1000{\rm GeV}$ and $\upsilon_{L_t}=3000{\rm GeV}$.

\begin{figure}[h]
\setlength{\unitlength}{1mm}
\centering
\includegraphics[width=3.3in]{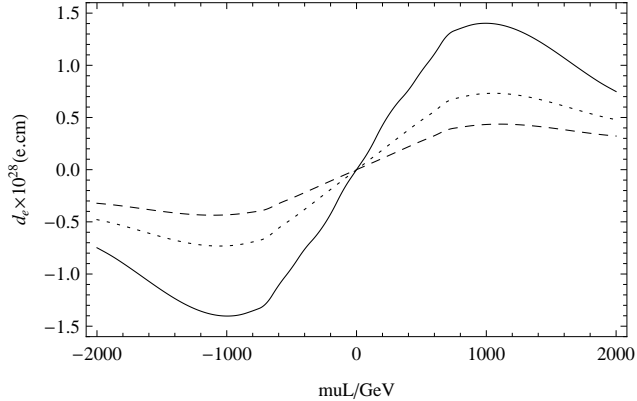}
\caption[]{With $\theta_{\mu_L}=0.5\pi$ and $ \theta_1=\theta_2=\theta_{\mu}=\theta_L=0$, the contributions to electron EDM varying with $muL$
are plotted by the solid line, dotted line and dashed line respectively corresponding to $S_R=S_L=(6\times10^6,8\times10^6,10\times10^6){\rm GeV}^2$.}\label{dEMUL}
\end{figure}

Supposing $ \theta_1=\theta_2=\theta_{\mu}=\theta_L=0$, we study the contributions from $\theta_{\mu_L}$ to electron EDM. 
$\mu_L$ relates with
sneutrino mass squared matrix and lepton neutralino mass  matrix. Here, the contributions to lepton EDM from the
lepton neutralino-slepton diagram are dominant, because the chargino-sneutrino diagram contributions are suppressed by
the tiny neutrino Yukawa couplings through $\texttt{Im}[(Y_\nu)_{IJ}^*(Z_{\nu^{IJ}})^{2i}(Z_{\nu^{IJ}})^{1i*}]$. With
$\theta_{\mu_L}=0.5\pi, \tan\beta=15, \mu_H=-800{\rm GeV}, m_2=800{\rm GeV}, m_1=1000{\rm GeV}$, in Fig.(\ref{dEMUL})
we plot the solid line, dotted line and dashed line versus $muL$ $(-2000\sim2000~{\rm GeV})$  corresponding to $S_R=S_L=(6\times10^6,8\times10^6,10\times10^6){\rm GeV}^2$.
When $|muL|$ is around 1000GeV, $|d_e|$ reaches its biggest value. For the same positive $muL$, the solid line is up the dotted line and the dotted line
is up the dashed line. It implies that heavier slepton masses lead to smaller lepton neutralino-slepton contributions.
The largest values of the three lines are respectively $1.4\times10^{-28}e.cm, 0.8\times10^{-28}e.cm$ and $0.4\times10^{-28}e.cm$.
As $|muL|>1000$GeV, $|d_e|$ is the decreasing function of $|muL|$, which is reasonable because large $muL$ should suppress the results.

\begin{figure}[h]
\setlength{\unitlength}{1mm}
\centering
\includegraphics[width=3.3in]{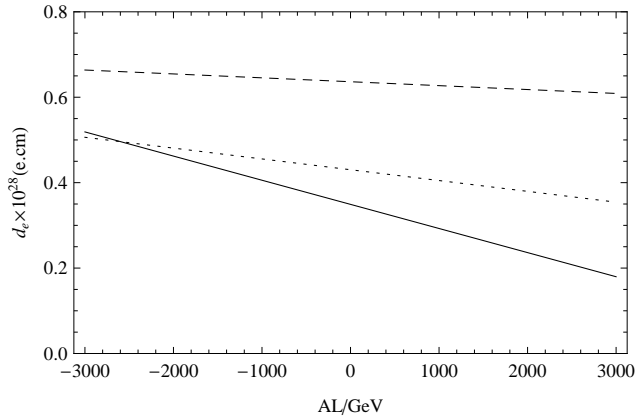}
\caption[]{With $\theta_\mu=0.5\pi, \theta_1=\theta_2=\theta_{\mu_L}=\theta_L=0$ and $\tan\beta=(10,15,25)$, the contributions to electron EDM varying with $AL$
are plotted by the solid line, dotted line and dashed line respectively.}\label{dEAL}
\end{figure}

In the follow of this subsection, the parameters
$S_L=6.0\times10^6 {\rm GeV}^2, S_R=1.0\times10^6 {\rm GeV}^2, \mu_L=-3000{\rm GeV}$ are adopted.
The mass squared matrix of slepton has the parameter $AL$ leading to the influence
of electron EDM. With the parameters
$m_{1}=1000{\rm GeV},~m_2=600{\rm GeV},MU=-800{\rm GeV},\theta_\mu=0.5\pi,
 (A'_l)_{11}=(A'_l)_{22}=(A'_l)_{33}=200{\rm GeV}$ and $\tan\beta=(10,15,25)$,
the results are shown by the solid line, dotted line and dashed line respectively in Fig.(\ref{dEAL}).
These three lines are all decreasing functions of $AL$ in the region $(-3000\sim3000)$GeV, and their values vary
in the range $(1.8\sim6.6)\times10^{-29}e.cm$. Generally speaking, larger $\tan\beta$ leads to larger $d_e$.
The contributions from the right-handed sneutrino have nothing to do with $AL$. For the dashed line,
the right-handed sneutrino contributions are about $1.4\times10^{-29}e.cm$; For the dotted line,
the right-handed sneutrino corrections are around $1.0\times10^{-29}e.cm$; For the solid line,
the right-handed sneutrino contributions are about $0.7\times10^{-29}e.cm$. The lepton neutralino-slepton diagram also
gives important contributions and they vary from $0.5 \times10^{-29}e.cm$ to $1.5 \times10^{-29}e.cm$ for the
three lines. It is obviously that the contributions from the right-handed sneutrino and lepton neutralino are very important.

\begin{figure}[h]
\setlength{\unitlength}{1mm}
\centering
\includegraphics[width=3.3in]{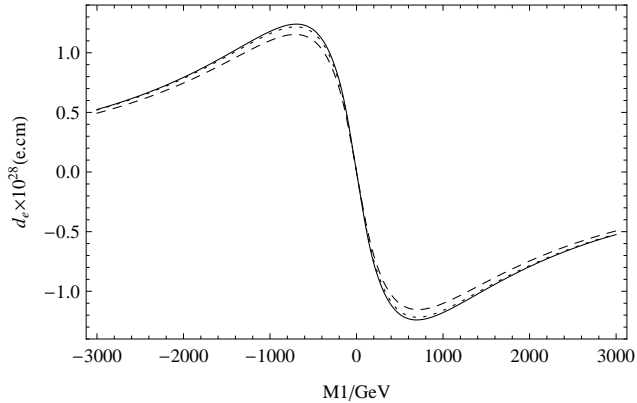}
\caption[]{With $\theta_1=0.5\pi, \theta_2=\theta_\mu=\theta_{\mu_L}=\theta_L=0$ and $\mu_H=(-500,-1000,-2000)$GeV, the contributions to electron EDM varying with $M1$
are plotted by the solid line, dotted line and dashed line respectively.}\label{dEM1}
\end{figure}

$m_1$ relates with the neutralino mass matrix. So we study the numerical results versus $M1$ with
 $\theta_1=0.5\pi,m_2=750{\rm GeV},(A'_l)_{11}=-135{\rm GeV},
 (A'_l)_{22}=(A'_l)_{33}=200{\rm GeV},\tan\beta=15,
AL=-2000{\rm GeV},\mu_H=(-500,-1000,-2000){\rm GeV}$, and the corresponding results
are plotted by the solid line, dotted line and dashed line in Fig.(\ref{dEM1}). The three
lines are very similar. The biggest value is about $12\times10^{-29}e.cm$, as $M1=-600$GeV.
The absolute value of $d_e$ turns small slowly with the enlarging $|M1|$ in the
region$(600\sim3000)$GeV. When $|M1|$ is biggish, the masses of neutralinos are
heavy, which suppresses the contributions to electron EDM. At the point $M1=0$,
there is none CP violating effect and $d_e=0$
is reasonable. The right-handed sneutrino contributions are related with $\theta_2$ and $\theta_\mu$ through
the coupling with chargino. The lepton neutralino-slepton contributions have relations with $\theta_{\mu},\theta_{\mu_L}, \theta_L$.
In this condition, only $\theta_1$ is nonzero, then both the right-handed sneutrino and lepton neutralino
give none contribution to $d_e$.

\begin{figure}[h]
\setlength{\unitlength}{1mm}
\centering
\includegraphics[width=3.3in]{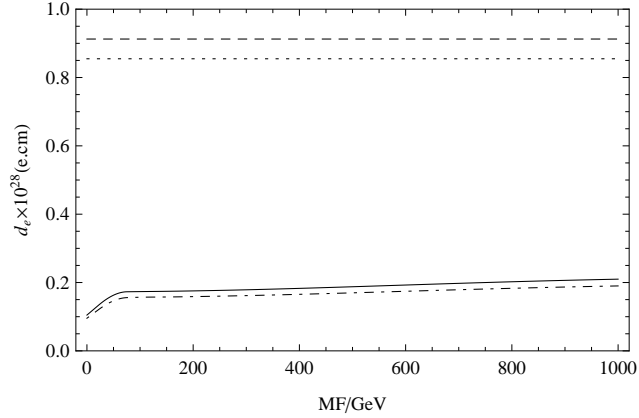}
\caption[]{With $\theta_2=0.5\pi, \theta_1=\theta_\mu=\theta_{\mu_L}=\theta_L=0$ and $M2=(700,1000)$GeV, the contributions to electron EDM varying with $MF$
are plotted by the dashed line and dotted line respectively; the right-handed sneutrino  contributions
are plotted by the solid line and dot-dashed line respectively.}\label{dESNUFDJ}
\end{figure}

$m_2$ is included in the mass matrices of neutralino and chargino, so $\theta_2$ is very important for  electron EDM.
As $\theta_2=0.5\pi$, we study the effects from some non-diagonal elements of $m^2_{\tilde{N}^c}$. With the
parameters $m_{1}=1000{\rm GeV},\mu_H=-800{\rm GeV},(A'_l)_{11}=(A'_l)_{22}=(A'_l)_{33}=200{\rm GeV},\tan\beta=15,
 AL=-2000{\rm GeV}$, we adopt $(m^2_{\tilde{N}^c})_{12}=(m^2_{\tilde{N}^c})_{21}=MF^2$, and
 the other non-diagonal elements of $m^2_{\tilde{N}^c}$ are zero. For $M2=(700,1000){\rm GeV}$,
 the total one loop results are represented by the dashed line and dotted line respectively in the Fig.(\ref{dESNUFDJ}).
 At the same time, in Fig.(\ref{dESNUFDJ}) the contributions from the right-handed sneutrino are also plotted by the solid line and
 dot-dashed line corresponding to $M2=(700,1000){\rm GeV}$. The dashed line and dotted line are in the region $(8.5\sim 9.3)\times 10^{-29}e.cm$
  and they are both very slowly increasing functions of $MF$. The solid line and dot-dashed line increase quickly, when $MF$ turns
  from 0 to 60 GeV. In the $MF$ region $(60\sim1000)$GeV, the solid line and dot-dashed line turn large slowly. As $MF=0$,
  the right-handed sneutrino corrections are around $1.0\times 10^{-29}e.cm$. When $MF$ is larger than
  60 GeV, the contributions from the right-handed sneutrino can reach $2.0\times 10^{-29}e.cm$.
   Therefore, they
  are important and decrease the effects from the left-handed sneutrino to some extent with  nonzero $MF$.
    Because of $\theta_\mu=\theta_{\mu_L}=\theta_L=0$, lepton neutralino-slepton diagram does not give corrections to electron EDM in this condition.
 From the Figs.(\ref{dEMUL},  \ref{dEAL},
 \ref{dEM1}, \ref{dESNUFDJ}), one can find the upper bound of electron EDM
 is strict and has rigorous bound on the parameter space.

\subsection{the muon EDM}
 Lepton EDM is CP violating which is generated by the
CP violating phases. In the similar way, the muon EDM is numerically studied.
The parameters $S_L=7.0\times10^6 {\rm GeV^2},S_R=6.0\times10^6 {\rm GeV^2}$,
$S_\nu=2.0\times10^6{\rm GeV^2},AL=-1000{\rm GeV},\mu_L=-3000{\rm GeV},
ML=2000{\rm GeV},(A'_l)_{11}=(A'_l)_{22}=(A'_l)_{33}=200{\rm GeV}$ are supposed here.

\begin{figure}[h]
\setlength{\unitlength}{1mm}
\centering
\includegraphics[width=3.3in]{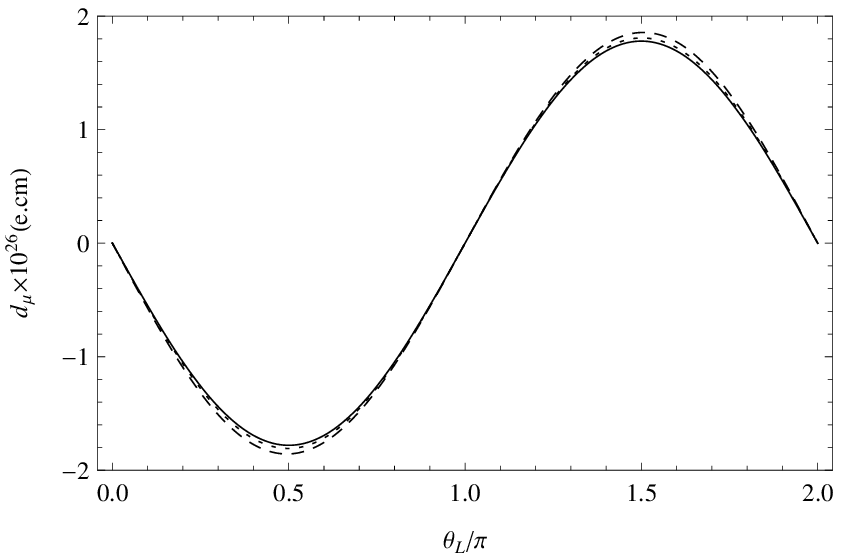}
\caption[]{With $\theta_1=\theta_2=\theta_\mu=\theta_{\mu_L}=0$ and $\upsilon_{L_t}=(1,3,5){\rm TeV}$,
the contributions to muon EDM varying with $\theta_L$
are plotted by the solid line, dotted line and dashed line respectively.}\label{dMML}
\end{figure}

The lepton neutrino mass matrix includes the new CP violating phase $\theta_L$.
Therefore, it obviously produces new contributions to the lepton EDM.
As $\theta_1=\theta_2=\theta_{\mu}=\theta_{\mu_L}=0$, $m_1=700{\rm GeV},m_2=800{\rm GeV},\mu_H=-600{\rm GeV}, \tan\beta=15$,
in Fig.(\ref{dMML}) we study $d_\mu$ versus $\theta_L$ with $\upsilon_{L_t}=(1,3,5){\rm TeV}$, and the results are plotted by the
solid line, dotted line and dashed line. The three lines are of the same shape, and all look like $-\sin\theta_L$.
These lines are almost coincident, whose  largest values are about $1.9\times 10^{-26}(e.cm)$.
The Fig.(\ref{dMML}) implies the effects from $\upsilon_{L_t}$ to muon EDM are small.
These contributions are only come from
lepton neutralino-slepton diagram. The chargino-sneutrino and netralino-slepton diagrams give zero contribution to $d_\mu$.

\begin{figure}[h]
\setlength{\unitlength}{1mm}
\centering
\includegraphics[width=3.3in]{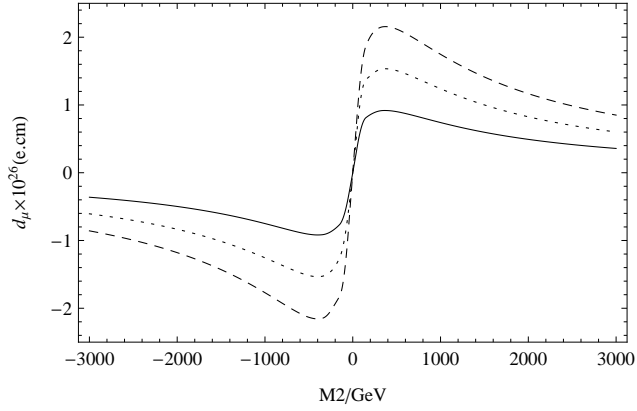}
\caption[]{With $\theta_2=0.5\pi, \theta_1=\theta_\mu=\theta_{\mu_L}=\theta_L=0$ and $\tan\beta=(15,25,35)$, the contributions to muon EDM varying with $M2$
are plotted by the solid line, dotted line and dashed line respectively.}\label{dMM2}
\end{figure}

As noted in the front subsection, $m_2$ is an important parameter for the lepton EDM.
Using $m_{1}=700{\rm GeV},\theta_2=0.5\pi,\mu_H=-600{\rm GeV},\upsilon_{L_t}=3000{\rm GeV},
\tan\beta=(15,25,35)$, we study
$d_\mu$ versus $M2$ in Fig.(\ref{dMM2}), and the numerical results are plotted by the solid line, dotted line and dashed line respectively.
At the point $M2=\pm400$ GeV, the absolute value of each line reaches its biggest value.
The dashed line can arrive at $2.0\times10^{-26}e.cm$. Larger $|M2|$ leads to smaller $d_\mu$ for the three lines, when $|M2|$
is larger than 400GeV. As $|M2|$ is very big, heavy neutralinos and charginos are produced, which suppresses the contributions to muon EDM.
The order from big to small for the absolute values of the three lines is the dashed line $>$ the dotted line $>$
the solid line. The corrections from the right-handed sneutrino are about $(23\sim25)\%$ of $d_\mu$.
As $\theta_\mu=\theta_{\mu_L}=\theta_L=0$, lepton neutralino-slepton contributions are zero.

\begin{figure}[h]
\setlength{\unitlength}{1mm}
\centering
\includegraphics[width=3.3in]{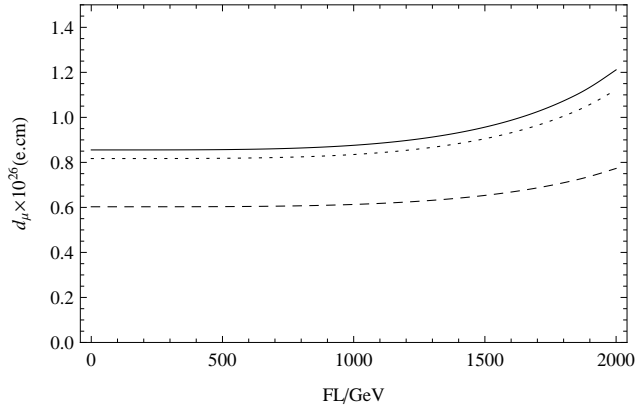}
\caption[]{With $\theta_2=0.5\pi, \theta_1=\theta_\mu=\theta_{\mu_L}=\theta_L=0$ and $MU=(-500,-700,-1500)$GeV, the contributions to muon EDM varying with $FL$
are plotted by the solid line, dotted line and dashed line respectively.}\label{dMMLF}
\end{figure}

$m_{\tilde{L}}^2$ and $m_{\tilde{R}}^2$ influence the masses of slepton and sneutrino.
The non-diagonal elements of $m_{\tilde{L}}^2$ and $m_{\tilde{R}}^2$ may give considerable contributions.
To simplify the discussion, we suppose the relations
$(m_{\tilde{L}}^2)_{ij}=(m_{\tilde{R}}^2)_{ij}=FL^2$, for $(i\neq j; ~i,j=1,2,3)$ and adopt the
parameters $m_{1}=1000{\rm GeV},M2=750{\rm GeV},\upsilon_{L_t}=3000{\rm GeV},\theta_2=0.5\pi,
\tan\beta=15$. In Fig.(\ref{dMMLF}),
the solid line, dotted line and dashed line, respectively represent the results with
$\mu_H=(-500,-700,-1500){\rm GeV}$. They are all increasing functions of $FL$, and the solid line is the highest line.
The dotted line is the middle one. These numerical results are in the range $0.6\sim1.2\times10^{-26}e.cm$.
When $FL$ varies from 0 to 50GeV, though the total results do not have significant change,
the contributions of the right-handed sneutrino increase quickly. The reason is that left-handed sneutrino
corrections turn small fast with the non-diagonal elements of $m_{\tilde{L}}^2$ and $m_{\tilde{R}}^2$.
Right-handed sneutrino and left-handed sneutrino mix and
should be regarded as an entirety. To some extent, non-zero $FL$ moves the contributions from left-handed sneutrino
to the right-handed sneutrino, without affecting the total results obviously.
The ratios for the right-handed sneutrino contributions to $d_{\mu}$ ran from $20\%$ to bigger values, and
can even reach $50\%$.
With our used parameters, the numerical results for muon EDM shown as the Figs.
(\ref{dMML},\ref{dMM2},\ref{dMMLF}) are
about at the order of  $10^{-26}e.cm$, which is almost seven-order smaller than muon EDM upper bound.

\subsection{the tau EDM}
Tau is the heaviest lepton, whose EDM upper bound is the largest one and at the order of $10^{-17}e.cm$. Tau EDM is
also of interest and calculated here. In this subsection, we use
 $S_\nu=2.0\times10^6{\rm GeV^2},AL=-1000{\rm GeV}$,$(A'_l)_{11}=(A'_l)_{22}=(A'_l)_{33}=200{\rm GeV},m_{1}=1000{\rm GeV},
 \mu_L=-3000{\rm GeV},ML=3000{\rm GeV},\upsilon_{L_t}=3000{\rm GeV}$.

\begin{figure}[h]
\setlength{\unitlength}{1mm}
\centering
\includegraphics[width=3.3in]{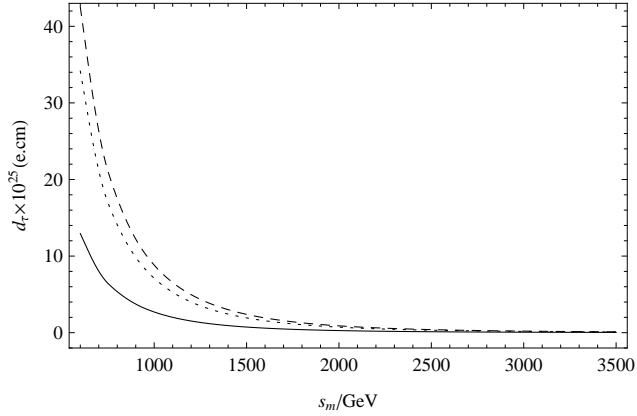}
\caption[]{With $\theta_1=\theta_2=\theta_\mu=\theta_{\mu_L}=0$ and $\theta_L=(-0.1,-0.3,-0.5)\pi$,
 the contributions to tau EDM varying with $s_m$
are plotted by the solid line, dotted line and dashed line respectively.}\label{dTMLS}
\end{figure}

With $\theta_1=\theta_2=\theta_\mu=\theta_{\mu_L}=0, \tan\beta=15, m_2=800{\rm GeV}, \mu_H=-800{\rm GeV}$ and supposing $S_L=S_R=s_m^2$, we plot the results versus $s_m$ by the
solid line, dotted line and dashed line for $\theta_L=(-0.1,-0.3,-0.5)\pi$. Under this supposition, only lepton neutralino-slepton diagram
gives nonzero corrections.
In Fig.(\ref{dTMLS}), one can find
$d_\tau$ is the decreasing function of $s_m$. In the $s_m$ region $(600 \sim 1000)$GeV, the three lines decrease quickly with the
enlarging $s_m$.  As $s_m=600{\rm GeV}$, the dashed line can reach  $4\times10^{-24}e.cm$ and even larger.
When $s_m>1000 {\rm GeV}$, the results decrease slowly and are almost coincident as $s_m>2000 {\rm GeV}$.
At the points $s_m=(1000,2000){\rm GeV}$, the results are at the order of $(10^{-25},10^{-26})e.cm$.

\pagebreak[4]

\begin{figure}[h]
\setlength{\unitlength}{1mm}
\centering
\includegraphics[width=3.3in]{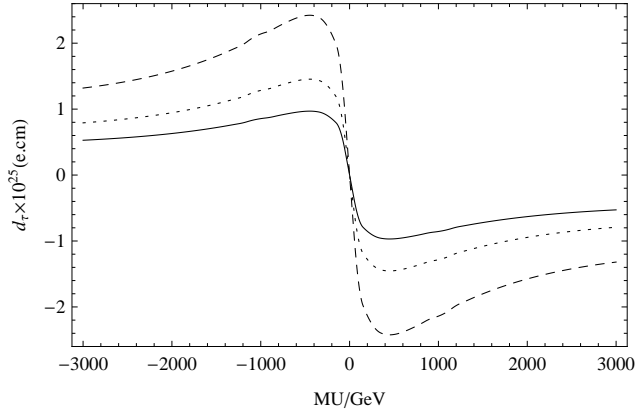}
\caption[]{With $\theta_\mu=0.5\pi, \theta_1=\theta_2=\theta_{\mu_L}=\theta_L=0$ and $\tan\beta=(10,15,25)$, the contributions to tau EDM varying with $MU$
are plotted by the solid line, dotted line and dashed line respectively.}\label{dTMU}
\end{figure}

The relation between $d_\tau$ and $\mu_H$ is studied here. We use the parameters
$S_L=7.0\times10^6 {\rm GeV^2}, S_R=6.0\times10^6 {\rm GeV^2}$, $m_2=750{\rm GeV},\theta_\mu=0.5\pi,\tan\beta=(10,15,25)$, and plot the results by
the solid line, dotted line and dashed line respectively in the Fig.(\ref{dTMU}). The dashed line reaches $2.4\times10^{-25}e.cm$ as $MU=-500{\rm GeV}$.
When $|MU|>500{\rm GeV}$, the abstract values of the numerical results  shrink with the enlarging $|MU|$.
Larger $\tan\beta$ results in larger $d_\tau$, when the other parameters are same. The right-handed
sneutrino contributions are about $(15\sim20)\%$ of the total results for the three lines. At the same time, the lepton neutralino-slepton
contributions can match the right-handed sneutrino contributions, and they are at the same order.

\begin{figure}[h]
\setlength{\unitlength}{1mm}
\centering
\includegraphics[width=3.3in]{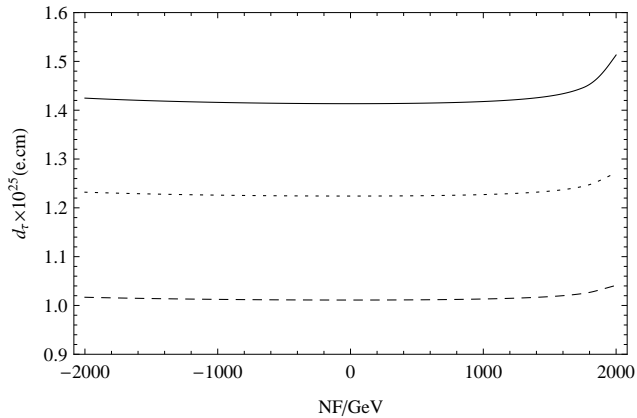}
\caption[]{With $\theta_2=0.5\pi, \theta_1=\theta_\mu=\theta_{\mu_L}=\theta_L=0$ and $M2=(500,1000,1500)$GeV, the contributions to tau EDM varying with $NF$
are plotted by the solid line, dotted line and dashed line respectively.}\label{dTANF}
\end{figure}

The parameters in BLMSSM and absent in MSSM should affect  $d_\tau$ to some extent.
With the parameters $S_L=7.0\times10^6 {\rm GeV^2}, S_R=6.0\times10^6 {\rm GeV^2}$, $\theta_2=0.5\pi,\mu_H=-800{\rm GeV},
\tan\beta=15$ and based on the assumption $(A_{N^c})_{ij}=(A_{N})_{ij}=NF$,
for $i\neq j$ and $i,j=1,2,3$, we research the effects from the non-diagonal elements of $A_{N^c}$ and $A_{N}$.
In Fig.(\ref{dTANF}), the solid line, dotted line and dashed line correspond respectively to the results obtained with
$M2=(500,1000,1500){\rm GeV}$. When $NF$ is near 2000GeV, the results increase observably. The solid line is up the
dotted line, and the dotted line is up the dashed line. The values of the three lines are in the region $(1.0\sim1.5)\times10^{-25}e.cm$.
The right-handed sneutrino contributions largen quickly with the increasing $|NF|$, and their ratios to the total results
improve from $24\%$ to $50\%$. The reason should be that the non-diagonal elements of $A_{N^c}$ and $A_{N}$
weaken the left-handed sneutrino contributions. At the same time the contributions from the right-hand sneutrino
are enhanced. Generally speaking, in BLMSSM the left-and right-handed sneutrinos are an integral whole, and should be discussed
together.
The Figs.(\ref{dTMLS},\ref{dTMU},\ref{dTANF})
show that the one-loop contributions to tau EDM are at the order of $10^{-25}\sim10^{-24}(e.cm)$ in our used parameter space. These contributions are about eight-order smaller than the upper bound of tau EDM.

\section{discussion and conclusion}
   In the frame work of CP violating BLMSSM, we study the one-loop contributions to the lepton$(e,\mu,\tau)$ EDM. The used parameters
can satisfy the experiment data of Higgs and neutrino.
The effects of
the CP violating phases $\theta_1,\theta_2,\theta_\mu,\theta_{\mu_L},\theta_L$ to the lepton EDM are researched. The upper bound of electron EDM is $8.7\times10^{-29}e.cm$, which
gives strict confine on the BLMSSM parameter space. In our used parameter space, the contributions to electron EDM can easily
reach its upper bound and even exceed it. The numerical values obtained for muon EDM and tau EDM are at the order of
$10^{-26}e.cm$ and $10^{-25}\sim10^{-24}e.cm$ respectively. They both are several-order smaller than their EDM upper bounds.
 Our numerical results
mainly obey the rule $d_e/d_\mu/d_\tau\sim m_e/m_\mu/m_\tau$. The right-handed sneutrino contributions are considerable, and should be taken into account. The contributions from the lepton neutralino-slepton have two new CP violating sources, and include the coupling constant $g_L$. If we enlarge $g_L$ and adopt other parameters, the lepton neutralino-slepton
contributions to lepton EDM can enhance several orders.
In general, the numerical results of the
lepton EDM are large, and they maybe detected by the experiments in the near future.

{\bf Acknowledgements}

   This work has been supported by the National Natural
Science Foundation of China (NNSFC) with Grants
No. (11275036, 11447111), the open project of
State Key Laboratory of Mathematics-Mechanization with
Grant No. Y5KF131CJ1, the Natural Science Foundation
of Hebei province with Grant No. A2013201277, and the Found of Hebei province with
the Grant NO. BR2-201 and the
Natural Science Fund of Hebei University with Grants
No. 2011JQ05 and No. 2012-242, Hebei Key Lab of Optic-Electronic Information and Materials, the midwest universities comprehensive strength promotion project.

\end{document}